\newif\ifarxiv
\arxivtrue
\ifarxiv
  \documentclass[sigconf,nonacm]{acmart}
\else
  \documentclass[sigconf,review,anonymous]{acmart}
\fi

\AtBeginDocument{%
  }

\usepackage{enumitem}
\usepackage{amsmath}

\usepackage{xcolor}
\usepackage{colortbl}
\usepackage[most]{tcolorbox}
\usepackage{makecell}

\usepackage{multirow}
\usepackage{subcaption}
\usepackage{xspace}

\makeatletter
\newcommand{\rmnum}[1]{\romannumeral #1}
\newcommand{\Rmnum}[1]{\expandafter\@slowromancap\romannumeral #1@}
\makeatother

\newcommand{\ie}{\emph{i.e., }}
\newcommand{\eg}{\emph{e.g., }}

\definecolor{ourcolor1}{RGB}{70, 120, 180}

\definecolor{ourcolor}{RGB}{225, 238, 250}

\newcommand{\ours}{\mbox{TS-Rec}\xspace}

\newcommand{\slh}[1]{{\color{black}{#1}}}

\def \ustc {$^{\hspace{.1em}{\color{blue}\boldsymbol{u}}}$}
\def \nus {$^{\hspace{.1em}{\color{purple}\boldsymbol{n}}}$}
\def \whu {$^{\hspace{.1em}{\color{green}\boldsymbol{w}}}$}

\setcopyright{acmlicensed}
\copyrightyear{2018}
\acmYear{2018}
\acmDOI{XXXXXXX.XXXXXXX}

\acmConference[Conference acronym 'XX]{Make sure to enter the correct
  conference title from your rights confirmation email}{June 03--05,
  2018}{Woodstock, NY}
\acmISBN{978-1-4503-XXXX-X/2018/06}

\makeatletter
\def\blfootnote{\gdef\@thefnmark{}\@footnotetext}
\makeatother

\begin{document}

\title{Fine-grained Semantics Integration for Large Language Model-based Recommendation}

\author{Jiawei Feng\ustc, Xiaoyu Kong\ustc, Leheng Sheng\nus, Bin Wu\whu, Chao Yi \\ Feifang Yang, Xiang-Rong Sheng, Han Zhu, Xiang Wang\ustc, Jiancan Wu$^{\dagger}$\ustc, Xiangnan He$^{\dagger}$\ustc} 
    \email{{jwf3ng, xiangwang1223, wujcan, xiangnanhe}@gmail.com} 
    \email{{kongxy}@mail.ustc.edu.com}
    \email{{leheng.sheng}@u.nus.edu}
    % \email{{wubin2021}@whu.edu.cn}
    
    \affiliation{
        \institution{\ustc University of Science and Technology of China, Hefei, Anhui, China}
        \country{}
    }
    \affiliation{
        \institution{\nus National University of Singapore, Singapore}
        \country{}
    }
    \affiliation{
        \institution{\whu Wuhan University, Wuhan, Huibei, China}
        \country{}
    }

\renewcommand{\shortauthors}{Feng et al.}

\begin{abstract}
    % Recent advances in Large Language Models (LLMs) have shifted in recommendation systems from the discriminative paradigm to the LLM-based generative paradigm, where the recommender autoregressively generates sequences of semantic identifiers (SIDs) for target items conditioned on historical interaction.
    % While prevalent LLM-based recommenders have demonstrated performance gains by aligning pretrained LLMs between the language space and the SID space, modeling the SID space still faces two fundamental challenges: (\rmnum{1}) Semantically Meaningless Initialization: SID tokens are randomly initialized, severing the semantic linkage between the SID space and the pretrained language space at start point, and (\rmnum{2}) Coarse-grained Alignment: existing SFT-based alignment tasks primarily focus on item-level optimization, while overlooking the semantics of individual tokens within SID sequences.
    Recent advances in Large Language Models (LLMs) have driven a shift in recommender systems from the discriminative paradigm to the LLM-based generative paradigm, where the recommender autoregressively generates sequences of semantic identifiers (SIDs) for target items conditioned on historical interaction.
    While prevalent LLM-based recommenders have demonstrated performance gains by aligning pretrained LLMs between the language space and the SID space, modeling the SID space still faces two fundamental challenges: (\rmnum{1}) Semantically Meaningless Initialization: SID tokens are initialized by random sampling without incorporating token-specific semantics, leaving them without explicit semantic correspondence to the pretrained language space; and (\rmnum{2}) Coarse-grained Alignment: existing SFT-based alignment tasks primarily focus on item-level optimization while overlooking the semantics of individual tokens within SID sequences.
    
    To address these challenges, we propose \textbf{\ours}, which can integrate \textbf{T}oken-level \textbf{S}emantics into LLM-based \textbf{Rec}ommenders. Specifically, \ours comprises two key components:
    (1) Semantic-Aware Embedding Initialization (SA-Init), which initializes SID token embeddings by applying mean pooling to the pretrained embeddings of keywords extracted by a teacher model; and
    (2) Token-level Semantic Alignment (TS-Align), which aligns individual tokens within the SID sequence with the shared semantics of the corresponding item clusters.
    Extensive experiments on three real-world benchmarks demonstrate that \ours consistently outperforms traditional and generative baselines across all standard metrics.
    % The results demonstrate that integrating fine-grained semantic information consistently enhances the performance of LLM-based generative recommenders.
    These findings confirm the effectiveness of incorporating fine-grained semantic information into LLM-based generative recommenders.
    The implementation is available at \url{https://anonymous.4open.science/r/TS-Rec-Review-8F99/}.

% \blfootnote{$^\dagger$Corresponding Author.}
\end{abstract}




% \received{20 February 2007}
% \received[revised]{12 March 2009}
% \received[accepted]{5 June 2009}

\maketitle
\begin{figure}[!t]
\centering
\includegraphics[width=0.48\textwidth]{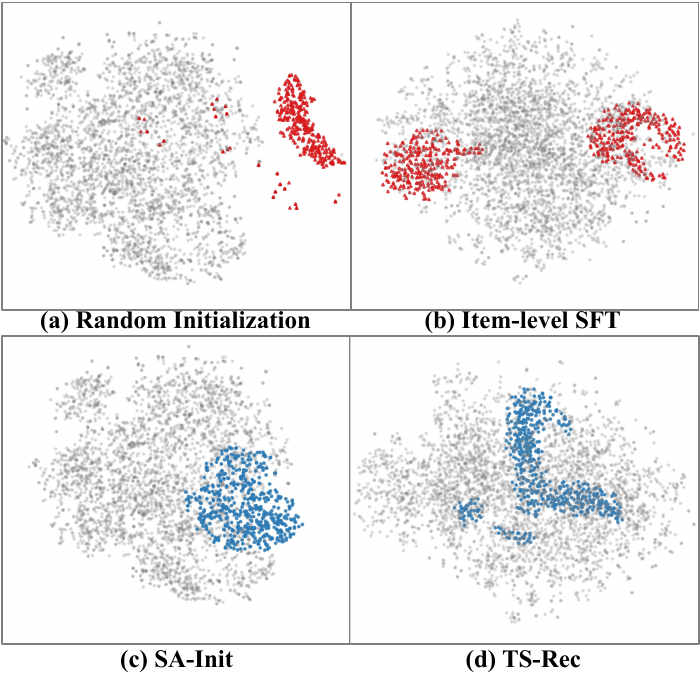}
\vspace{-25pt}
% \caption{Representations of SIDs before (\ie left column) and after (\ie right column) alignment. The randomly initialized SIDs are initially far away from the semantic space of the LLM (\ie grey points), while SA-Init enables SIDs within the semantic space of the LLM. This makes the SID alignment of the randomly initialized one more difficult than SA-Init.}
\caption{Representations of SIDs before (\ie left column) and after (\ie right column) alignment. The randomly initialized SIDs are initially far away from the semantic space of the LLM (\ie grey points), while SA-Init enables SIDs within the semantic space of the LLM. Consequently, aligning randomly initialized SID embeddings is more difficult than aligning embeddings produced by SA-Init.}
\label{fig:tsne}
\end{figure}

\section{Introduction}

\slh{Large language model (LLM)-based generative recommendation has recently demonstrated great potential in modeling user preferences~\citep{LCREC_DBLP:conf/icde/ZhengHLCZCW24,TALLREC_DBLP:conf/recsys/BaoZZWF023,LLaRA_DBLP:conf/sigir/LiaoL0WYW024, ALPHARec_DBLP:journals/corr/abs-2407-05441,LLM4RecSurvey_DBLP:journals/www/WuZQWGSQZZLXC24}. 
In generative recommendation, items are represented as discrete tokens rather than continuous vectors\citep{TIGER_DBLP:conf/nips/RajputMSKVHHT0S23,LETTER_DBLP:conf/cikm/0007BLZ0FNC24, VQRec_DBLP:conf/www/HouHMZ23,HSTU_DBLP:conf/icml/ZhaiLLWLCGGGHLS24}, enabling the model to directly generate the token corresponding to the next item through an autoregressive process. 
Building on this paradigm, LLM-based generative recommendation further incorporates large language models to leverage their rich world knowledge, enhancing the recommender’s understanding of items and thereby improving recommendation performance\citep{ONERECTHINK_DBLP:journals/corr/abs-2510-11639,MTGR_DBLP:conf/cikm/HanYCJJ0MHLJHZY25,DBLP:journals/corr/abs-2512-24762, CoLLM}.}

\begin{figure}[t]
\centering
\includegraphics[width=\linewidth]{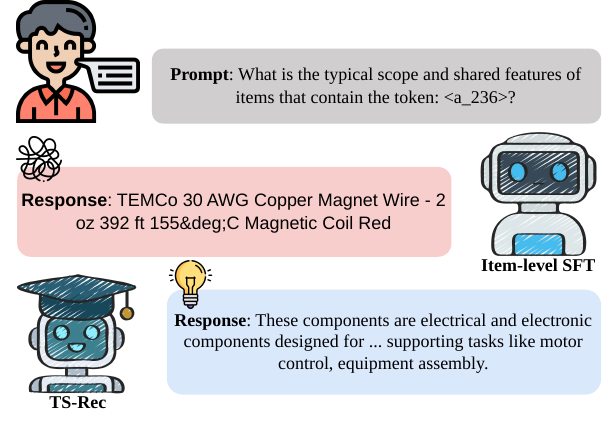}
\vspace{-20pt}
\caption{\slh{TS-Rec enables recommenders to have high-level and fine-grained understanding of token-level SIDs.}}
\label{fig:teaser}
\end{figure}

\slh{
In a typical LLM-based generative recommendation pipeline~\citep{ONERECTHINK_DBLP:journals/corr/abs-2510-11639,DBLP:journals/corr/abs-2510-24431,DBLP:journals/corr/abs-2505-16994}, the training procedure can be broadly divided into three stages:
(1) item tokenization as semantic identifiers (SIDs)\citep{TIGER_DBLP:conf/nips/RajputMSKVHHT0S23,LETTER_DBLP:conf/cikm/0007BLZ0FNC24,VQRec_DBLP:conf/www/HouHMZ23}: each item is first represented as a sequence of $N$ fixed-length tokens (\ie SIDs) that lie outside the original vocabulary of the LLM (\eg three tokens like <a\_236> <b\_231> <c\_226>), commonly obtained using vector quantization methods such as RQ-VAE\citep{zeghidour2021soundstream};
(2) alignment\citep{LCREC_DBLP:conf/icde/ZhengHLCZCW24,LETTER_DBLP:conf/cikm/0007BLZ0FNC24} between these out-of-vocabulary (OOV) item tokens and the LLM backbone through supervised fine-tuning (SFT), enabling the model to understand these tokens for recommendation;
and (3) preference learning (\eg reinforcement learning (RL)\citep{DBLP:journals/corr/abs-2510-12211,DBLP:journals/corr/abs-2510-24431,DBLP:journals/corr/abs-2512-24762,DBLP:journals/corr/abs-2506-13695,DBLP:journals/corr/abs-2510-23077,DBLP:journals/corr/abs-2505-16994} or direct preference optimization (DPO)~\citep{DBLP:conf/nips/ChenTZ0SZWC24,DMPO,DBLP:conf/nips/RafailovSMMEF23,DBLP:journals/corr/abs-2511-11255}) to further enhance the model's ability to capture user preferences, thereby pushing the upper bound of recommendation quality.
}
\slh{
As the first two steps, item tokenization and alignment with LLMs directly determine how the LLM understands items and thus largely affect recommendation performance, playing vital roles in LLM-based generative recommendation. 
However, existing approaches~\citep{TIGER_DBLP:conf/nips/RajputMSKVHHT0S23, LCREC_DBLP:conf/icde/ZhengHLCZCW24, DBLP:journals/corr/abs-2510-24431} for these two steps remain coarse, which limits the LLM’s ability to effectively understand OOV item tokens and consequently leads to suboptimal recommendation performance.
Specifically, we identify two key limitations in current item tokenization and alignment with LLMs as follows:}
\begin{itemize}[leftmargin=*]
    % \item \slh{$\textbf{Semantically Meaningless Initialization.}$ 
    % % In the item tokenization stage, SIDs are typically initialized in a semantically meaningless manner, deviating from the LLM representation space and hindering their subsequent alignment with LLMs. 
    % \fjw{In the tokenization stage of items}, SIDs are typically initialized in a semantically meaningless manner, deviating from the LLM representation space and hindering their subsequent alignment with LLMs. 
    % As illustrated in Figure \ref{fig:tsne}(a), the current SID initialization strategy relies on random initialization, resulting in representations that deviate substantially from the intrinsic representation space of the LLM. 
    % Consequently, aligning these SIDs to the LLM becomes more challenging, as they must be pulled from a distant region of the representation space (\cf Figure \ref{fig:tsne}(a)).}
    \item \slh{$\textbf{Semantically Meaningless Initialization.}$
    After item tokenization, newly introduced SID tokens are typically initialized by random sampling from a distribution fitted to the pretrained token embeddings. Although this strategy approximately matches the global statistics of the pretrained embedding space, it does not incorporate the semantics of the item cluster represented by each SID token. Consequently, the initialized SID embeddings lack explicit semantic correspondence with language tokens, leaving the model to establish such associations entirely during subsequent SFT. As illustrated in Figure~\ref{fig:tsne}(a), this semantic mismatch produces a clear distributional gap between the initialized SID representations and semantically meaningful language representations.}
    \item  \slh{$\textbf{Coarse-grained Alignment.}$ 
    During the alignment stage with the LLM, current alignment methods tend to adopt a coarse-grained manner~\citep{TIGER_DBLP:conf/nips/RajputMSKVHHT0S23, LCREC_DBLP:conf/icde/ZhengHLCZCW24, DBLP:journals/corr/abs-2510-24431}, where the alignment task is constructed to translate the SID into natural language (\eg translate <a\_236> <b\_231> <c\_226> into \textit{``Relay/Capacitor Hard Start Kit''}). 
    However, this alignment strategy is relatively coarse and lacks a fine-grained understanding of SIDs at the token level. 
    As illustrated in Figure~\ref{fig:teaser}, when we ask the LLM what general feature the first token <a\_236> represents, it instead responds with a specific item, which is unsatisfactory.
    In practice, however, a token at the SID level typically corresponds to a broader category or attribute, rather than an individual item. This mismatch indicates that the LLM fails to capture fine-grained semantics at the token level, resulting in insufficient token-level understanding and further hindering the recommendation performance.
    }
\end{itemize}

\slh{We argue that the fine-grained semantics integration is crucial in both item tokenization and alignment with LLM stages, while current approaches largely ignore such token-level semantics.
To address the above limitations, we introduce \ours, a generative recommendation framework that enhances SID comprehension, which comprises two key components:
}
\begin{itemize}[leftmargin=*]
    \item $\textbf{Semantic-Aware Embedding Initialization.}$ We initialize SID token embeddings in a semantically aware manner by mean-pooling the pretrained embeddings of keyword sequences extracted by a semantic extractor. Specifically, for each token, the semantic extractor summarizes the common properties shared among all items containing that token, generating a corresponding token description and keywords. We then initialize the SID token embeddings via the mean pooling of the pre-trained embeddings of keyword sequences. This ensures that each SID token is endowed with meaningful semantics prior to post-alignment, thus providing a better starting point for the subsequent process.
    \item $\textbf{Token-level Semantic Alignment.}$ Building upon standard SFT objectives, we formulate a token-level alignment task to align the SID space and the language space at a finer granularity. Specifically, we use the token descriptions generated by the semantic extractor and their corresponding SID tokens as training pairs. This alignment objective ensures that the model captures both the semantics of individual SID tokens and complete SID sequences. By grounding these token-wise semantic units in the LLM's language space, the alignment strengthens the LLM's fine-grained understanding of SIDs and promotes their effective use in the generation process.
\end{itemize}

To evaluate the performance of \ours, we conduct extensive experiments on three real-world Amazon benchmarks: Industrial, Office, and Toys~\citep{Amazon_ni2019justifying}. Empirical results demonstrate that \ours consistently outperforms the strongest baselines across all reported metrics, yielding relative improvements of up to 9.76\%, 9.45\%, and 13.07\% on Industrial, Office, and Toys, respectively.
Moreover, our analysis reveals that the superiority of \ours stems from its ability to bridge the semantic gap between discrete semantic identifiers and the LLM's language space, providing a more robust optimization starting point and a more granular understanding of SIDs. Furthermore, we show that \ours serves as a superior semantic backbone that significantly elevates the performance ceiling of the subsequent "SFT-then-RL" pipeline~\citep{DBLP:journals/corr/abs-2510-24431}. These findings underscore the critical importance of fine-grained semantic integration for LLM-based recommendation.

\section{Related Work}

% 总起一下，判别式向生成式转变，两个研究问题：1）SID 2）利用LLM的世界知识
Traditional recommendation systems have been dominated by discriminative paradigms, such as collaborative filtering~\citep{DBLP:conf/www/HeLZNHC17,DBLP:conf/www/SarwarKKR01} and click-through rate (CTR) models~\citep{DBLP:conf/kdd/ZhouZSFZMYJLG18,DBLP:conf/aaai/ZhouMFPBZZG19,MUSE_DBLP:journals/corr/abs-2512-07216, SIMTIER_DBLP:conf/cikm/ShengYGWCZCZG0J24}. 
Recently, the field has experienced a paradigm shift toward \textit{generative recommendation}~\citep{DBLP:journals/corr/abs-2511-10138, TIGER_DBLP:conf/nips/RajputMSKVHHT0S23, HSTU_DBLP:conf/icml/ZhaiLLWLCGGGHLS24, DBLP:journals/corr/abs-2506-13695, LCREC_DBLP:conf/icde/ZhengHLCZCW24, DBLP:journals/corr/abs-2512-24762, DBLP:journals/corr/abs-2512-22386, MTGR_DBLP:conf/cikm/HanYCJJ0MHLJHZY25}, which reframes recommendation as a next-item generation problem, where transformer-based models~\citep{QWEN2.5_DBLP:journals/corr/abs-2412-15115, T5_DBLP:journals/jmlr/RaffelSRLNMZLL20} autoregressively generate recommended items conditioned on user interaction histories. This paradigm offers greater generalizability and significantly improves performance in cold-start and long-tail scenarios.

% 仅基于SID做GR，没有利用LLM的语言建模能力和世界知识
\subsection{SID-based Generative Recommendation}
A cornerstone of the generative paradigm is the design of Semantic Identifiers (SIDs), which serve as discrete tokens bridging continuous item multimodal representations and discrete item identifiers. 

The pioneering work TIGER~\citep{TIGER_DBLP:conf/nips/RajputMSKVHHT0S23} utilizes residual quantization (RQ-VAE)~\citep{RQVAE_DBLP:conf/cvpr/LeeKKCH22} to derive hierarchical SIDs from item metadata.
Building upon this foundation, more effective codebook construction strategies are explored to obtain higher-quality SIDs~\citep{QARM_DBLP:conf/cikm/LuoCSYHYLZ0HQZZ25, PSRQ_DBLP:conf/cikm/WangOXWRXGL25,ONESEARCH_DBLP:journals/corr/abs-2509-03236}.
Subsequent studies including LETTER~\citep{LETTER_DBLP:conf/cikm/0007BLZ0FNC24}, EAGER~\citep{EAGER_DBLP:conf/kdd/Wang0GW0S25}, and UNGER~\citep{UNGER_10.1145/3773771} have further integrated collaborative signals into the generative process.
In industrial settings, HSTU~\citep{HSTU_DBLP:conf/icml/ZhaiLLWLCGGGHLS24} optimizes the sequence modeling efficiency through a hierarchical transformer architecture tailored for high-throughput recommendation. OneRec~\citep{DBLP:journals/corr/abs-2506-13695} further establishes an end-to-end training pipeline to effectively deploy generative recommendation in large-scale industrial systems.
% 不足：最新的生成式模型的潜力未被充分探索

Despite these advancements, traditional SID-based generative methods only encode item semantics into the randomly initialized SID embeddings, while largely overlooking the potential of \textbf{exploiting the inherent semantic capabilities} of the generative backbone model.

% 早期基于title ID，后与SID结合，但因为SID随机初始化无法充分利用LLM本身的能力，浪费资源在对齐上
\subsection{LLM-based Generative Recommendation}
Another line of work focuses on utilizing modern LLMs~\citep{QWEN2.5_DBLP:journals/corr/abs-2412-15115, LLAMA3_DBLP:journals/corr/abs-2407-21783} as the recommendation backbone to harness their vast semantic understanding and world knowledge~\citep{TALLREC_DBLP:conf/recsys/BaoZZWF023, LLaRA_DBLP:conf/sigir/LiaoL0WYW024, InteraRec_DBLP:conf/pakdd/KarraT24, ONERECTHINK_DBLP:journals/corr/abs-2510-11639,DBLP:journals/corr/abs-2512-24762,ilora}. 

Early efforts like TallRec~\citep{TALLREC_DBLP:conf/recsys/BaoZZWF023} and LLaRa~\citep{LLaRA_DBLP:conf/sigir/LiaoL0WYW024} demonstrate the effectiveness of instruction tuning, employing item titles to align LLMs with recommendation objectives via specialized prompts. However, relying solely on textual titles to represent items has notable limitations: Long or blurry titles often strain the LLM's context window and exacerbate the risk of hallucinations. To address this, LC-Rec~\citep{LCREC_DBLP:conf/icde/ZhengHLCZCW24} and OneRec-Think~\citep{ONERECTHINK_DBLP:journals/corr/abs-2510-11639,DBLP:journals/corr/abs-2512-24762} adopt compact SID-based representations for items, and align the representation spaces of pre-trained LLMs and SIDs through token expansion and multi-task supervised fine-tuning. Building on these SFT-aligned models, reinforcement learning (RL) algorithms like GRPO~\citep{GRPO_DBLP:journals/corr/abs-2402-03300} are further introduced to refine recommendation quality~\citep{DBLP:journals/corr/abs-2512-24762, DBLP:journals/corr/abs-2510-24431,DBLP:journals/corr/abs-2510-12211}.

Initializing newly added vocabulary tokens from existing pretrained embeddings is a common strategy for vocabulary expansion. Such methods typically exploit lexical or morphological relationships between a new token and existing sub-tokens. However, SID tokens are artificial discrete codes with no linguistic surface forms, making these relationships unavailable. Our SA-Init therefore derives token-specific semantics from the item cluster represented by each SID token and maps the extracted keywords into the pretrained embedding space. The distinction lies in the source of the aggregated semantics rather than in the mean-pooling operation itself. Specifically, SA-Init uses cluster-level item characteristics instead of the linguistic form of the new token.

Nevertheless, existing approaches typically initialize newly added SID tokens randomly and align LLMs only at the level of complete hierarchical SIDs, overlooking \textbf{fine-grained semantic learning at the individual token level}. Since RL optimization relies on the SFT model as its starting point, insufficient SID understanding at this stage can limit subsequent gains. These limitations hinder effective SID comprehension and motivate our work.

\begin{figure*}[!t]
% \vspace{-0.2cm}
\setlength{\abovecaptionskip}{0.3cm}
\setlength{\belowcaptionskip}{-0.0cm}
\centering
\includegraphics[width=0.9\linewidth]{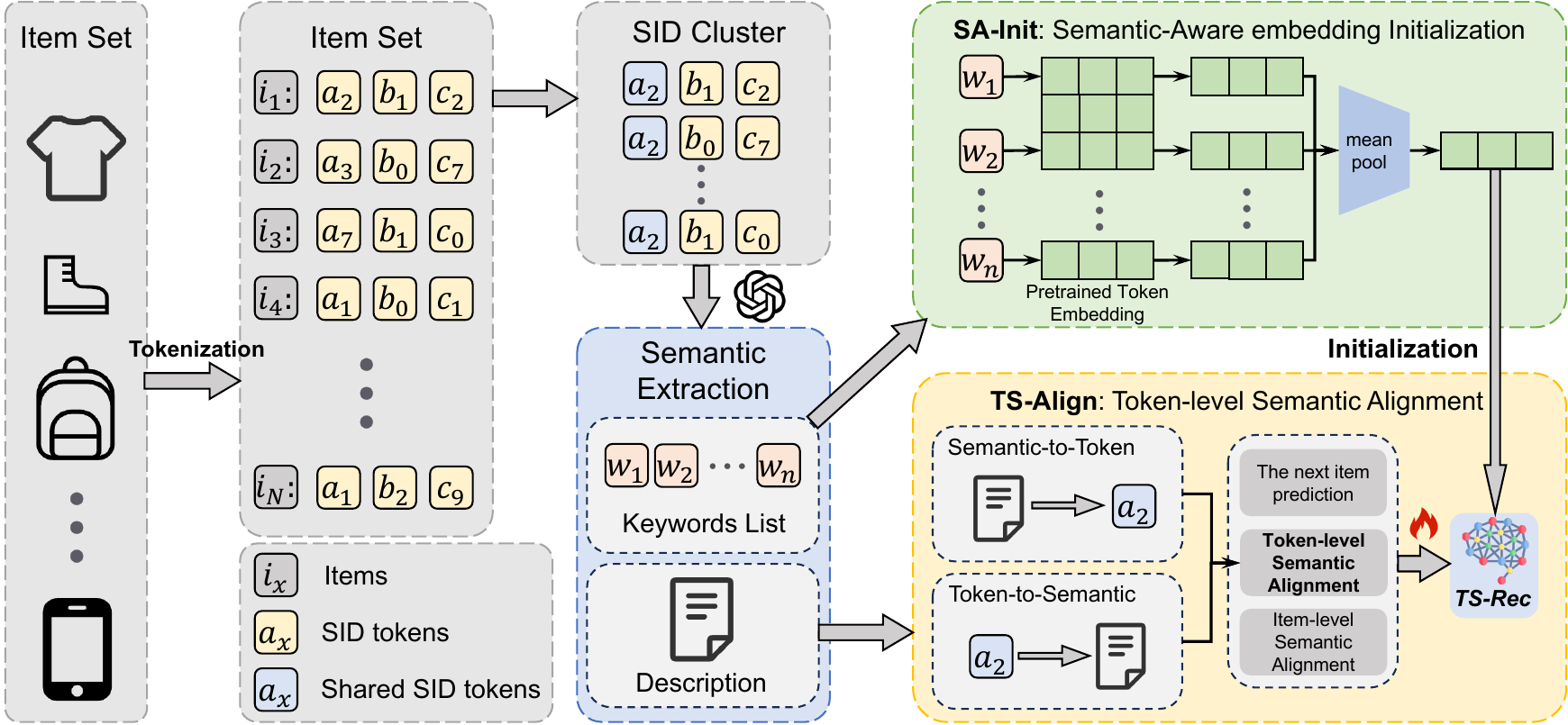}
\caption{The overall framework of our \ours. The pipeline consists of two complementary modules: (1) SA-Init, which initializes SID embeddings with semantic descriptions generated by a semantic extractor from clustered items; (2) TS-Align, which aligns each SID token with its semantics through multi-task instruction tuning.}
\label{fig:ts-rec}
\end{figure*}

\section{Preliminaries}

This section introduces the key concepts and notation used throughout the paper. Specifically, we formulate the generative recommendation task and describe the tokenization mechanism and SFT objective.

\subsection{Task Formulation}

Let $H_{u}$ be the chronologically ordered interaction history of user $u$. Each item $i\in H_{u}$ is represented as a three-level SID tuple $e_{i}=\{s^{(1)}_{i}, s^{(2)}_{i}, s^{(3)}_{i}\}$, where $s^{(l)}_{i}$ is the SID token at level $l$. Given $H_{u}$, the generative recommender $\pi_{\theta}$, parameterized by $\theta$, is trained to predict the preferred item $i_{p}$ that matches the preference of user $u$ from the item set.

\subsection{Tokenization and Embedding Initialization}\label{sec:tokenization_and_emb}

% In this section, we introduce how items are tokenized into hierarchical semantic identifiers (SIDs) via RQ-KMeans~\citep{QARM_DBLP:conf/cikm/LuoCSYHYLZ0HQZZ25} and how to incorporate these out-of-vocabulary tokens into a pre-trained LLM.

\subsubsection{Item Tokenization.} For each item $i$, we concatenate its textual title and description into a single sentence, which is then fed into a frozen encoder to produce a $d$-dimensional embedding $\mathbf{x}\in\mathbb{R}^{d}$. The continuous embedding is subsequently discretized using the RQ-KMeans algorithm~\citep{QARM_DBLP:conf/cikm/LuoCSYHYLZ0HQZZ25}. 

Let $\tilde{\mathbf{M}} = [\mathbf{x}_1; \dots; \mathbf{x}_N] \in \mathbb{R}^{N \times d}$ be the matrix stacking the embeddings of all $N$ items. We initialize $\mathbf{R}^{(1)} = \tilde{\mathbf{M}}$. For each layer $l \in \{1, \dots, L\}$, where $L$ is the number of hierarchical levels in the semantic codebook, we learn a codebook $\mathbf{C}^{(l)} = \{\mathbf{c}_k^{(l)}\}_{k=1}^{K_l}$ by running K-means with $K_l$ centroids on the current residuals $\mathbf{R}^{(l)}$:
\begin{equation}
\mathbf{C}^{(l)} = \text{K-means}(\mathbf{R}^{(l)}, K_l).
\end{equation}
For item $i$ ($1 \le i \le N$), the nearest centroid index is assigned as:
\begin{equation}
s_i^{(l)} = \arg \min_k \|\mathbf{R}_i^{(l)} - \mathbf{c}_k^{(l)}\|_2,
\end{equation}
where $\|\cdot\|$ denotes the Euclidean norm. The residual is updated as:
\begin{equation}
\mathbf{R}_i^{(l+1)} = \mathbf{R}_i^{(l)} - \mathbf{c}_{s_i^{(l)}}^{(l)}.
\end{equation}

We set the number of codebook levels to $L=3$. Each item is thus represented by a sequence of semantic identifiers $\{s^{(1)}_{i}, s^{(2)}_{i}, s^{(3)}_{i}\}$, which serves as a unique\footnote{To avoid collisions, we reassign the last SID token randomly for any conflicting items.} token sequence for each item $i$ and will be used by the recommender model in autoregressive generation.

\subsubsection{Embedding Initialization.} Following the tokenization process, the newly generated SID tokens in $\mathcal{V}_{\text{SID}}$ are integrated into the pre-trained vocabulary of the LLM, denoted as $\mathcal{V}_{\text{Pre}}$,  to form an expanded vocabulary:
\begin{equation}
    \mathcal{V} = \mathcal{V}_{\text{SID}}\cup\mathcal{V}_{\text{Pre}}.
\end{equation}
% An embedding vector is then assigned to each newly added token in $\mathcal{V}_{\text{SID}}$. These embeddings are typically initialized by sampling from a Gaussian distribution $\mathcal{N}(\mu, \Sigma)$ fitted to the pre-trained token embeddings. Here, $\mu$ and $\Sigma$ denote the mean vector and covariance matrix of the LLM’s pre-trained embedding space.
An embedding vector is then assigned to each newly added token in $\mathcal{V}_{\text{SID}}$. These embeddings are typically initialized by random sampling from a Gaussian distribution $\mathcal{N}(\mu, \Sigma)$ fitted to the pre-trained token embeddings. Here, $\mu$ and $\Sigma$ denote the mean vector and covariance matrix of the LLM's pre-trained embedding space. Although this strategy matches the global statistics of the pre-trained embedding space, it does not provide token-specific semantic grounding for individual SID tokens.

\subsection{Supervised Fine Tuning for LLM-based Recommenders}

In supervised fine-tuning (SFT), the semantic understanding and generative capabilities of large language models (LLMs) are leveraged to enhance sequential recommendation. As detailed in Appendix~\ref{appendix:sft_tasks}, the fine-tuning tasks can be conveniently formulated as conditional language generation in a sequence-to-sequence manner. LLM-based recommenders are fine-tuned using the following language modeling objective:
\begin{equation}\label{eq:sft_objective}
    \mathcal{L}_{\text{SFT}}=-\sum_{(X,Y)}\sum^{|Y|}_{j=1}\log \pi_{\theta}(Y_{j}\mid X, Y_{<j}),
\end{equation}
where $(X,Y)$ denotes a pair of instruction and target response from the training data, $Y_{j}$ denotes the $j$-th token in the response $Y$, and $Y_{<j}$ denotes the prefix tokens preceding $Y_{j}$. The recommender $\pi_{\theta}$ models the conditional probability of generating the target response $Y$ given instruction $X$.

% \begin{figure*}[!t]
% % \vspace{-0.2cm}
% \setlength{\abovecaptionskip}{0.0cm}
% \setlength{\belowcaptionskip}{-0.0cm}
% \centering
% \includegraphics[scale=1.2]{figure/overview_1.pdf}
% \caption{The overall framework of our \ours. The pipeline consists of two complementary modules: (1) SA-Init, which initializes SID embeddings with semantic descriptions generated by an Semantic Extractor from clustered items; (2) TS-Align, which aligns each SID token with its semantics through multi-task instruction tuning.}
% \label{fig:ts-rec}
% \end{figure*}

\section{Method}

As illustrated in Figure~\ref{fig:ts-rec}, we propose a framework to bridge the semantic gap between SIDs and pre-trained LLMs. The method comprises two complementary modules: 
1) \textbf{SA-Init} (Sec.~\ref{method:sa-init}), which statically injects semantic priors into SIDs using an Extractor LLM to address the cold-start issue; 
and 2) \textbf{TS-Align} (Sec.~\ref{method:ts-align}), which dynamically refines these representations through multi-task instruction tuning.

\subsection{SA-Init: Semantic-Aware Embedding Initialization}\label{method:sa-init}

As discussed in Section~\ref{sec:tokenization_and_emb}, unlike natural language tokens, the SID tokens in $\mathcal{V}_{\text{SID}}$ are randomly initialized and lack inherent semantic meaning at the onset of training. This semantic gap often leads to suboptimal convergence. To bridge this gap, we introduce \textbf{Semantic-Aware Embedding Initialization (SA-Init)}. By employing an Extractor LLM $\mathcal{M}_{ext}$ to synthesize the common semantics associated with each SID token, SA-Init injects semantic priors into the SID tokens, ensuring they are semantically grounded before the fine-tuning stage.
Specifically, the SA-Init comprises the following steps: 

\begin{itemize}[leftmargin=*]

    \item \textbf{Token-Specific Item Clusters.} 
    We first derive the SID vocabulary $\mathcal{V}_{\text{SID}}$ from the global item set $\mathcal{I}$. For each unique SID token $s \in \mathcal{V}_{\text{SID}}$, we aggregate the set of items that contain this token at any level of their identifier hierarchy. Formally, the cluster for token $s$ is defined as:
    \begin{equation}
        \mathcal{I}_{s} = \{i \in \mathcal{I} \mid s \in e_i \},
    \end{equation}
    where $e_i$ represents the SID tuple of item $i$. This step groups items based on shared structural codes, serving as a proxy for shared semantics.
    
    \item \textbf{Semantic Extraction.}
    To interpret the semantic rationale behind the structural cluster $\mathcal{I}_{s}$, we first randomly sample a subset of items $\hat{\mathcal{I}}_{s}$. We then prompt the Extractor LLM $\mathcal{M}_{\text{ext}}$ (e.g., DeepSeek) to generate a textual description $\mathcal{D}_{s}$ capturing the shared characteristics, along with a concise list of keywords $\mathcal{W}_{s}$:
    \begin{equation}
        \mathcal{D}_{s}, \mathcal{W}_{s} = \mathcal{M}_{\text{ext}}(\mathcal{P}_{\text{ext}}(\hat{\mathcal{I}}_{s})),
    \end{equation}
    where $\mathcal{P}_{\text{ext}}$ denotes the prompt template for semantic extraction (see Appendix~\ref{appendix:prompt} for details).
    
    \item \textbf{Initialization via Keyword Aggregation.} 
    Finally, we bridge the gap between the extracted symbolic keywords and the continuous embedding space of the recommender.
    We concatenate the keywords in $\mathcal{W}_{s}$ and project them into the \textit{backbone model's vocabulary} space via tokenization, yielding a token sequence $\mathcal{T}_{s}$:
    \begin{equation}
        \mathcal{T}_{s} = \text{Tok}_{\text{backbone}}(\text{Concat}(\mathcal{W}_{s})),
    \end{equation}
    where $\mathcal{T}_{s}$ consists of token IDs from the pre-trained vocabulary. To endow the SID token $s$ with an informative initial state, we perform an embedding lookup for each sub-token $v \in \mathcal{T}_{s}$ in the pre-trained embedding matrix $\mathbf{E}$. The embedding of $s$ is initialized by aggregating these semantic priors:
    \begin{equation}
        \mathbf{e}_{s} = \frac{1}{|\mathcal{T}_{s}|} \sum_{v \in \mathcal{T}_{s}} \mathbf{E}[v].
    \end{equation}
    This aggregation effectively aligns the initialized SID embeddings with the well-structured semantic space of the pre-trained LLM.
\end{itemize}

\subsection{TS-Align: Token-level Semantic Alignment}
\label{method:ts-align}

% --- Motivation ---
To equip the LLM with a fine-grained understanding of the SID structure, we introduce \textbf{Token-level Semantic Alignment (TS-Align)}.
Standard alignment approaches typically align the entire SID sequence of an item with its global description (e.g., title). However, this holistic alignment often overlooks the specific semantic contribution of individual tokens.
In contrast, TS-Align aligns each \textit{individual} SID token $s$ with its distilled semantic description $\mathcal{D}_s$ (derived in Section~\ref{method:sa-init}). This establishes an explicit mapping between the atomic structural units and their semantic meanings.

% --- Task Construction ---
Specifically, we construct a token-level instruction tuning dataset $\mathcal{D}_{\text{token}}$, consisting of bidirectional tasks formatted as instruction-response pairs $(x, y)$:

\begin{tcolorbox}[
colback=gray!5!white,
colframe=gray!60!black,
title=\textbf{Task 1: Semantic-to-Token Alignment (Understanding)},
fonttitle=\bfseries,
breakable,
boxrule=0.8pt,
left=2pt, right=2pt, top=2pt, bottom=2pt
]
\textbf{Instruction ($x$):}

Identify the specific SID token shared by items that exhibit the following scope and characteristics:
\textit{"$\strut\mathcal{D}_{s}$"}

\textbf{Response ($y$):}
$s$
\end{tcolorbox}

\begin{tcolorbox}[
colback=gray!5!white,
colframe=gray!60!black,
title=\textbf{Task 2: Token-to-Semantic Alignment (Generation)},
fonttitle=\bfseries,
breakable,
boxrule=0.8pt,
left=2pt, right=2pt, top=2pt, bottom=2pt
]
\textbf{Instruction ($x$):}

Describe the typical scope and shared features of items associated with the token: \textit{"$s$"}

\textbf{Response ($y$):}
$\mathcal{D}_{s}$
\end{tcolorbox}

% \noindent\textbf{Multi-Task Joint Optimization.}
% Following the multi-task learning paradigm established in generative recommendation \cite{LCREC_DBLP:conf/icde/ZhengHLCZCW24}, we treat the sequential recommendation as the primary task and the proposed TS-Align tasks as auxiliary sub-tasks.
% During the SFT phase, we jointly optimize the model on both the recommendation data $\mathcal{D}_{\text{rec}}$ and the token alignment data $\mathcal{D}_{\text{token}}$.
% By integrating these auxiliary alignment objectives, the model is regularized to maintain semantic consistency within the SID representations while minimizing the standard next-token prediction loss. This joint training ensures that the learned SIDs are not only distinct identifiers for recommendation but also semantically grounded representations.

\noindent\textbf{Multi-Task Joint Optimization.}
Following the multi-task learning paradigm established in generative recommendation \cite{LCREC_DBLP:conf/icde/ZhengHLCZCW24}, we treat the sequential recommendation as the primary task and the proposed TS-Align tasks as auxiliary sub-tasks.
During SFT, we construct the joint training set as
\begin{equation}
    \mathcal{D}
    =
    \mathcal{D}_{\mathrm{rec}}
    \cup
    \mathcal{D}_{\mathrm{align}}
    \cup
    \mathcal{D}_{\mathrm{token}},
\end{equation}
where $\mathcal{D}_{\mathrm{rec}}$ contains sequential recommendation instances, $\mathcal{D}_{\mathrm{align}}$ contains item-level SID--language alignment instances described in Appendix~\ref{appendix:sft_tasks}, and $\mathcal{D}_{\mathrm{token}}$ contains the two token-level alignment tasks introduced above. For $\mathcal{D}_{\mathrm{rec}}$, the instruction $X$ contains the interaction history and the response $Y$ is the SID sequence of the target item. For $\mathcal{D}_{\mathrm{align}}$, $(X,Y)$ forms an item-level SID--title pair, while for $\mathcal{D}_{\mathrm{token}}$, it forms a token-level SID--description pair.

We optimize the autoregressive language-modeling objective in Eq.~\eqref{eq:sft_objective} over the joint dataset $\mathcal{D}$. Training instances from the three task sets are mixed and shuffled within each mini-batch rather than optimized in alternating stages. Thus, no task-specific loss coefficient is introduced. The auxiliary token-level instances encourage the model to preserve semantic consistency within SID representations while learning the recommendation objective.
\section{Experiments}

In this section, we first describe our experimental setup, then present the results and analyze our approach to answer the following research questions:

\begin{itemize}[leftmargin=*]
\item \textbf{RQ1}: How does \ours perform compared to other baselines?

\item \textbf{RQ2}: How do the key components of \ours contribute to its recommendation performance?

\item \textbf{RQ3}: Does integrating reinforcement learning (RL)-based methods into \ours lead to greater performance improvements?

\item \textbf{RQ4}: Can \ours enhance the LLM's understanding of SIDs?

% \fjw{\item \textbf{RQ5}: Does \ours introduce any additional computational overhead?}
\end{itemize}

\subsection{Experimental Setting}

\begin{table}[htbp]
\centering
\caption{Statistics of the preprocessed datasets.}
\label{tab:data_statistics}
\begin{tabular}{lcccc}
\toprule
 \textbf{Datasets} & \textbf{\#Users} & \textbf{\#Items}  & \textbf{\#Interactions} & \textbf{Sparsity} \\
\midrule
Industrial & 7,694 & 3,685 & 45,324 & 99.84\% \\
Office & 8,328 & 3,459 & 48,656 & 99.83\% \\ 
Toys & 22,158 & 11,251 & 140,943 & 99.93\% \\
% Books & 24,812 & 15,532 & 280,988 & 99.93\% \\
\bottomrule
% \vspace{-10pt}
\end{tabular}
\end{table}

\subsubsection{Datasets.} We evaluate the performance of our proposed approach on three public subsets of the Amazon review dataset~\citep{Amazon_ni2019justifying}, specifically "Industrial and Scientific", "Office Products", and "Toys and Games". Each item is associated with a title and a description.
Following prior work\citep{LCREC_DBLP:conf/icde/ZhengHLCZCW24}, we filter out users and items with fewer than five interactions.
User interaction sequences are then constructed in chronological order.
The statistics of the preprocessed datasets are summarized in Table~\ref{tab:data_statistics}.

\begin{table*}[t]
\caption{Recommendation performance on three real-world datasets. HR and N denote Hit Ratio and NDCG, respectively. The best performance is highlighted in boldface, while the second-best performance is underlined.}
\vspace{-0.1in}
\label{tab:overall_performance}
\centering
\setlength{\tabcolsep}{1mm}{
    \resizebox{0.93\textwidth}{!} {
        \begin{tabular}{ll cccc cccc cccc}
        \toprule

        \multirow{2.4}{*}{Models} & \multicolumn{4}{c}{\textbf{Industrial}} & \multicolumn{4}{c}{\textbf{Office}} & \multicolumn{4}{c}{\textbf{Toys}} \\
        \cmidrule(lr){2-5} \cmidrule(lr){6-9} \cmidrule(lr){10-13} & HR@5 & N@5 & HR@10 & N@10 & HR@5 & N@5 & HR@10 & N@10 & HR@5 & N@5 & HR@10 & N@10 \\
        \midrule

        GRU4Rec & 0.0774 & 0.0598 & 0.0999 & 0.0669 & 0.0789 & 0.0595 & 0.1019 & 0.0669 & 0.0204 & 0.0143 & 0.0312 & 0.0178 \\
        Caser & 0.0717 & 0.0555 & 0.0942 & 0.0628 & 0.0865 & 0.0664 & 0.1093 & 0.0737 & 0.0280 & 0.0203 & 0.0397 & 0.0241 \\
        SASRec & 0.0909 & 0.0748 & 0.1088 & 0.0806 & 0.0949 & 0.0805 & 0.1120 & 0.0858 & 0.0431 & 0.0326 & 0.0581 & 0.0374 \\
        \midrule
        TIGER & 0.1010 & 0.0807 & 0.1321 & 0.0908 & 0.1163 & 0.0960 & \underline{0.1408} & 0.1002 & 0.0507 & 0.0356 & 0.0715 & 0.0423 \\
        HSTU & 0.1037 & \underline{0.0918} & 0.1163 & \underline{0.0958} & \underline{0.1252} & \underline{0.1079} & 0.1400 & \underline{0.1126} & 0.0621 & 0.0499 & 0.0705 & 0.0526 \\
        % LETTER & 0.0445 & 0.0294 & 0.0709 & 0.0378 & \underline{0.1315} & \underline{0.1074} & \underline{0.1520} & \underline{0.1139} & \underline{0.1080} & 0.0850 & \underline{0.1389} & \underline{0.0950} \\
        % LC-Rec & \underline{0.1057} & 0.0862 & \underline{0.1332} & 0.0952 & 0.1048 & 0.0859 & 0.1237 & 0.0920 & 0.0620 & 0.0463 & \underline{0.0860} & \underline{0.0539} \\
        LC-Rec & \underline{0.1057} & 0.0862 & \underline{0.1332} & 0.0952 & 0.1048 & 0.0859 & 0.1237 & 0.0920 & \underline{0.0686} & \underline{0.0503} & \underline{0.0906} & \underline{0.0574} \\

        % \midrule
        % ReaRec & 0.0568 & 0.0381 & 0.0843 & 0.0470 & 0.1173 & 0.0988 & 0.1385 & 0.1057 & 0.0973 & 0.0796 & 0.1205 & 0.0870 \\
        % R\textsuperscript{2}ec & \underline{0.0655} & \underline{0.0399} & \underline{0.0931} & \underline{0.0525} & 0.1147 & 0.0894 & 0.1486 & 0.1004 & 0.0880 & 0.0774 & 0.1253 & 0.0774 \\
        \midrule \rowcolor{ourcolor}
        \textbf{\ours} & \textbf{0.1153} & \textbf{0.0930} & \textbf{0.1462} & \textbf{0.1029} & \textbf{0.1307} & \textbf{0.1100} & \textbf{0.1541} & \textbf{0.1175} & \textbf{0.0756} & \textbf{0.0568} & \textbf{0.1008} & \textbf{0.0649} \\
        % \textbf{\ours} & \textbf{0.1153} & \textbf{0.0930} & \textbf{0.1462} & \textbf{0.1029} & \textbf{0.1307} & \textbf{0.1100} & \textbf{0.1541} & \textbf{0.1175} & \textbf{0.0793} & \textbf{0.0595} & \textbf{0.1051} & \textbf{0.0678} \\
        
        \bottomrule
        \end{tabular}
    }
}
% \vspace{-0.10in}
\end{table*}

\subsubsection{Evaluation Settings.} To evaluate performance, we adopt two widely used top-$K$ metrics: Hit Ratio (HR@$K$) and Normalized Discounted Cumulative Gain (NDCG@$K$). We report the main recommendation results at $K\in\{5,10\}$; for the RL experiments in RQ3, we additionally report results at $K=3$ following the evaluation setting of MiniOneRec. We perform full-ranking evaluation over the entire item set rather than sampled evaluation. For generative baselines that rely on beam search, the beam size is fixed at 20.

\subsubsection{Baselines.} We compare our approach with two categories of representative recommendation baselines: (1) traditional recommenders, including GRU4Rec, Caser, and SASRec; and (2) generative recommenders, including TIGER, HSTU, and LC-Rec.

The traditional recommendation models are as follows:
\begin{itemize}[leftmargin=*]
\item \textit{GRU4Rec}~\citep{GRU4REC_DBLP:conf/cikm/HidasiK18} is an RNN-based sequential recommendation model that uses gated recurrent units to capture users' evolving interests from their interaction sequences.
\item \textit{Caser}~\citep{CASER_DBLP:conf/wsdm/TangW18} is a CNN-based sequential recommendation model that applies convolutional filters to capture high-order patterns in recent user interactions.
\item \textit{SASRec}~\citep{SASREC_DBLP:conf/icdm/KangM18} is a widely used sequential recommendation model implemented with self-attention, enabling it to capture dynamic user preferences over time.
\end{itemize}

The generative recommendation models are as follows:
\begin{itemize}[leftmargin=*]
\item \textit{TIGER}~\citep{TIGER_DBLP:conf/nips/RajputMSKVHHT0S23} adopts a generative retrieval paradigm for recommendation and introduces Semantic IDs to index items.
\item \textit{HSTU}~\citep{HSTU_DBLP:conf/icml/ZhaiLLWLCGGGHLS24} employs a streaming architecture tailored for high-cardinality and non-stationary recommendation data.
\item \textit{LC-Rec}~\citep{LCREC_DBLP:conf/icde/ZhengHLCZCW24} is an LLM-based recommendation model that incorporates a series of tasks to align the SID space with the language space, thereby leveraging world knowledge to boost recommendation performance.
\end{itemize}

% \subsubsection{Baselines.} We compare proposed approach against two categories of leading recommendation baselines: (1) Traditional Recommenders: including GRU4Rec, Caser and SASRec. (2) Generative Recommenders: including TIGER ,HSTU and LC-Rec.  

\subsubsection{Implementation Details.}
To construct the item indices, we use the Qwen3-Embedding-4B encoder to generate continuous embeddings from item titles and descriptions. The codebook depth $L$ is set to 3; each layer contains 256 codebook vectors with a dimensionality of 32, optimized using the RQ-KMeans algorithm. To resolve index collisions, we randomly reassign the last SID token of each conflicting item.

For LLM fine-tuning, we implement our approach using Qwen2.5-1.5B~\citep{QWEN2.5_DBLP:journals/corr/abs-2412-15115}. The learning rate is set to $3\times10^{-4}$, with a total batch size of 1,024. We train the model for three epochs on each dataset. To prevent overfitting, we apply early stopping when the validation loss does not decrease for three consecutive evaluation steps. All experiments are conducted on eight NVIDIA H20 96GB GPUs.

\subsection{Overall Performance (RQ1)}

The experimental results of applying \ours to Qwen2.5-1.5B across three real-world benchmarks are presented in Table~\ref{tab:overall_performance}. We summarize our key observations as follows:

% (1) \textbf{\ours consistently achieves superior performance across all three benchmarks,} outperforming all compared baselines on every metric. Compared with the strongest baseline for each metric, \ours improves HR@5, NDCG@5, HR@10, and NDCG@10 by 9.08\%, 1.31\%, 9.76\%, and 7.41\%, respectively, on Industrial; by 4.39\%, 1.95\%, 9.45\%, and 4.35\% on Office; and by 27.70\%, 19.24\%, 22.21\%, and 25.79\% on Toys. The particularly large gains on Toys, the largest and sparsest dataset, indicate that fine-grained semantic integration remains effective in a more challenging recommendation setting. We attribute the improvements to two core factors: (a) Better Initialization: SA‑Init provides the LLM with semantic grounding at the token-embedding level, enabling more effective subsequent optimization; and (b) Fine-grained Semantic Alignment: TS‑Align preserves and leverages token-level semantics within SIDs during recommendation. These results validate fine-grained semantic integration as an effective approach for enhancing LLM-based recommenders.
(1) \textbf{\ours consistently achieves the best performance across all three datasets.} Compared with the strongest baseline for each metric, the relative improvements range from 1.31\% to 9.76\% on Industrial, from 1.95\% to 9.45\% on Office, and from 10.20\% to 13.07\% on Toys. The larger gains on Toys demonstrate the effectiveness of fine-grained semantic integration on a larger and sparser dataset. We attribute these gains to the semantically grounded initialization provided by SA-Init and token-level supervision from TS-Align.

% (2) \textbf{Generative recommendation models generally outperform traditional baselines across datasets.} This result highlights the effectiveness of autoregressive sequence modeling and semantic item representations for sequential recommendation. Among the generative baselines, LC-Rec further leverages pretrained language knowledge by aligning the language and SID spaces. Nevertheless, adapting LLMs to recommendation remains non-trivial because it requires aligning general-purpose language knowledge with domain-specific recommendation objectives while maintaining accurate item generation. By providing semantically grounded initialization and token-level alignment, \ours offers an effective way to address this challenge.

(2) \textbf{Generative recommenders generally outperform traditional sequential models.} This result demonstrates the effectiveness of autoregressive modeling and semantic item representations. Compared with existing generative methods, \ours further improves recommendation by explicitly grounding individual SID tokens in the LLM's language space.

% 5.57\% 3.61\% 9.08\% 1.30\% 23.27\% 7.41\%
% 5.82\% 2.42\% 4.39\% 1.94\% 9.44\% 4.35\%

% \begin{figure*}[!t]
%     \centering
%     \begin{subfigure}[b]{0.24\textwidth}
%         \includegraphics[width=\textwidth]{figure/hr_and_loss/Industrial_Loss_kddstyle.png}
%     \end{subfigure}
%     \begin{subfigure}[b]{0.24\textwidth}
%         \includegraphics[width=\textwidth]{figure/hr_and_loss/Industrial_HR5_kddstyle.png}
%     \end{subfigure}
%     \begin{subfigure}[b]{0.24\textwidth}
%         \includegraphics[width=\textwidth]{figure/hr_and_loss/Office_Loss_kddstyle.png}
%     \end{subfigure}
%     \begin{subfigure}[b]{0.24\textwidth}
%         \includegraphics[width=\textwidth]{figure/hr_and_loss/Office_HR5_kddstyle.png}
%     \end{subfigure}
%     % \vspace{-10pt}
%     \caption{Training dynamics of TS-Rec with different SA-Init depths on the \textit{Industrial} and \textit{Office} datasets. Configurations include \textit{Random Init}, \textit{SA-Init 1} (first layer), \textit{SA-Init 2} (first two layers), and \textit{SA-Init 3} (all three layers).}
%     \label{fig:hr_and_loss}
% \end{figure*}

\begin{figure*}[t]
    \centering
    \begin{subfigure}[b]{0.32\textwidth}
        \centering
        \includegraphics[width=\textwidth]{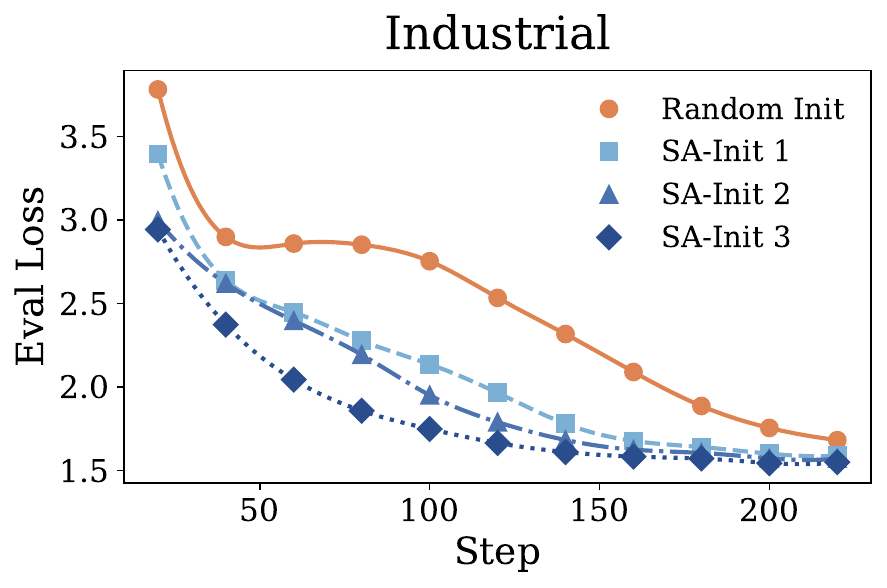}
    \end{subfigure}
    \hfill
    \begin{subfigure}[b]{0.32\textwidth}
        \centering
        \includegraphics[width=\textwidth]{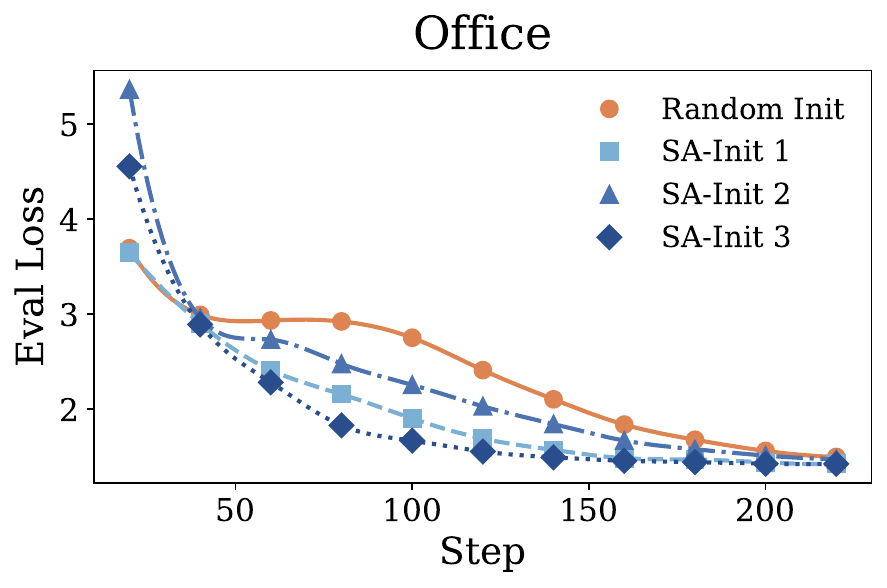}
    \end{subfigure}
    \hfill
    \begin{subfigure}[b]{0.32\textwidth}
        \centering
        \includegraphics[width=\textwidth]{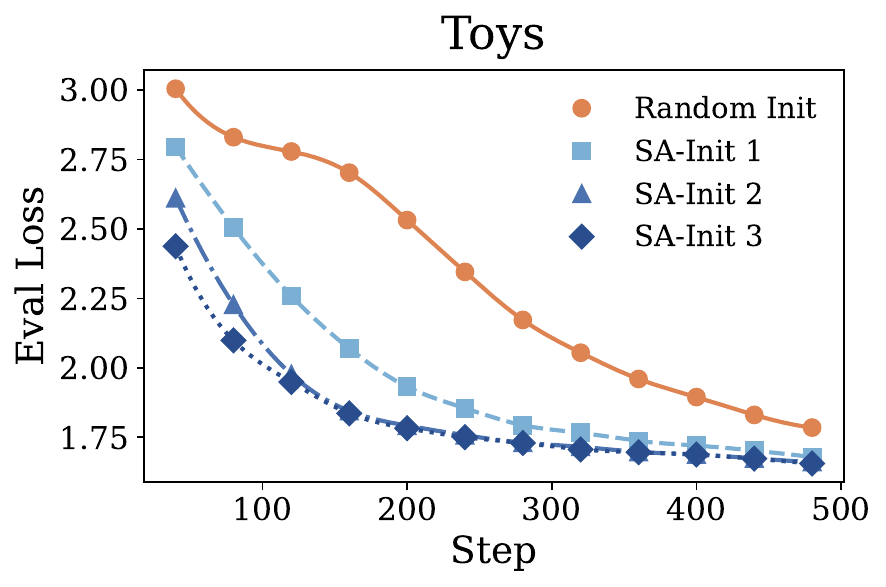}
    \end{subfigure}

    \vspace{0.3em}

    \begin{subfigure}[b]{0.32\textwidth}
        \centering
        \includegraphics[width=\textwidth]{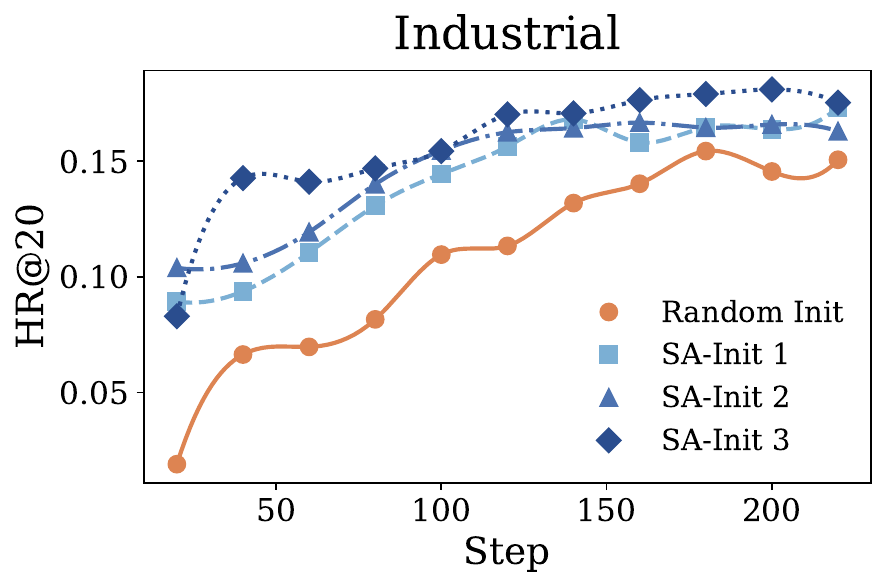}
    \end{subfigure}
    \hfill
    \begin{subfigure}[b]{0.32\textwidth}
        \centering
        \includegraphics[width=\textwidth]{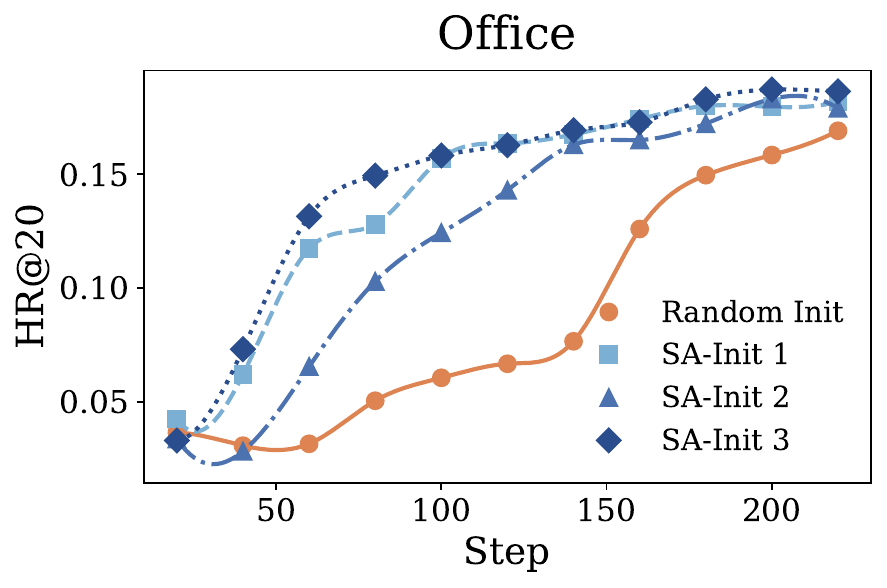}
    \end{subfigure}
    \hfill
    \begin{subfigure}[b]{0.32\textwidth}
        \centering
        \includegraphics[width=\textwidth]{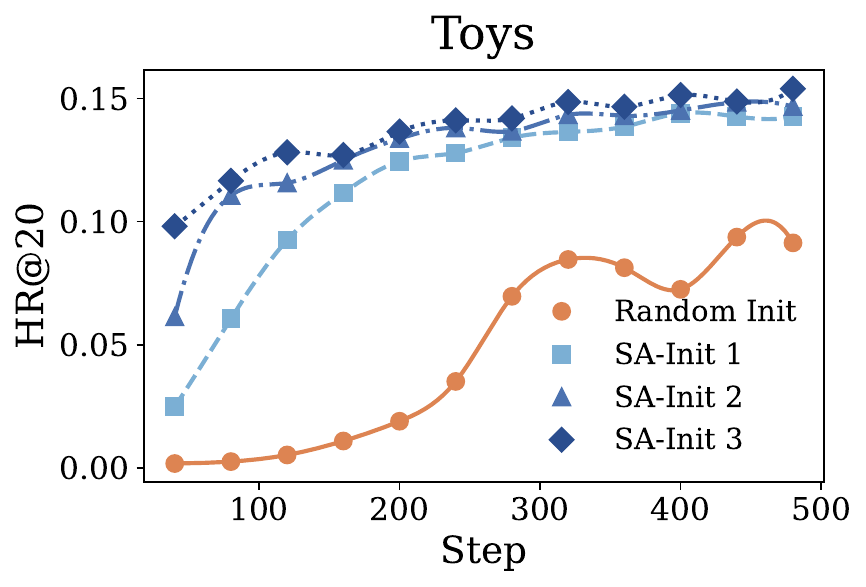}
    \end{subfigure}
    \vspace{-10pt}
    \caption{Training dynamics of \ours with different SA-Init depths on the \textit{Industrial}, \textit{Office}, and \textit{Toys} datasets. The top row reports evaluation loss, and the bottom row reports HR@20. Configurations include \textit{Random Init}, \textit{SA-Init 1} (first SID level), \textit{SA-Init 2} (first two SID levels), and \textit{SA-Init 3} (all three SID levels).}
    \label{fig:hr_and_loss}
\end{figure*}

\begin{table}[htbp]
\setlength{\tabcolsep}{3pt}
\caption{Ablation studies on the Industrial and Office datasets. HR@5 denotes Hit Ratio@5. The best performance is highlighted in boldface.}
\label{ablation}
% \vspace{-5pt}
\small
\begin{tabular}{lcccc}
\toprule
\multicolumn{1}{c}{\multirow{2}{*}{\textbf{Method}}} & \multicolumn{2}{c}{\textbf{Industrial}} & \multicolumn{2}{c}{\textbf{Office}} \\ \cmidrule(lr){2-3} \cmidrule(lr){4-5}
& \textbf{HR@5} & \textbf{NDCG@5} & \textbf{HR@5} & \textbf{NDCG@5} \\ \midrule \rowcolor{ourcolor}
\textbf{\ours} & \textbf{0.1153} & \textbf{0.0930} & \textbf{0.1307} & \textbf{0.1100} \\
\midrule
\quad w/o SA-Init & 0.1131 & 0.0911 & 0.1198 & 0.0950\\
\quad w/o TS-Align & 0.1111 & 0.0928 & 0.1300 & 0.1084\\
\quad w/o SA-Init \& TS-Align & 0.1080 & 0.0860 & 0.1178 & 0.1045\\ 
\bottomrule
\end{tabular}
\vspace{-5pt}
\end{table}

\subsection{Ablation Study (RQ2)}

In this section, we conduct ablation studies to validate the contribution of the core components in \ours. Our analysis focuses on two key aspects: (1) assessing the individual and combined impact of SA-Init and TS-Align, and (2) analyzing the influence of token initialization depth on the overall recommendation performance.

% \vspace{-5pt}

\subsubsection{Impact of SA-Init and TS-Align.}
To assess the individual contributions of SA-Init and TS-Align as well as their combined effect, we conduct controlled ablation experiments on the Industrial and Office datasets. The results are summarized in Table~\ref{ablation}, and the key findings are as follows:
(1) \textbf{Both SA-Init and TS-Align improve recommendation performance.} Removing either component consistently degrades HR@5 and NDCG@5. The effect of SA-Init is particularly pronounced on Office, where its removal decreases HR@5 and NDCG@5 by 8.34\% and 13.64\%, respectively.
(2) \textbf{SA-Init and TS-Align are complementary.} Removing both components yields the weakest performance on both datasets, confirming the benefit of combining semantically grounded initialization with token-level alignment.
% (1) \textbf{Both SA‑Init and TS‑Align contribute substantively to performance gains.} Removing SA‑Init results in notable performance degradation. On Industrial, HR@5 drops by 1.90\% (from 0.1153 to 0.1131) and NDCG@5 by 2.04\% (from 0.0930 to 0.0911). On Office, the drops are more pronounced: HR@5 decreases by 8.33\% (from 0.1307 to 0.1198) and NDCG@5 by 13.36\% (from 0.1100 to 0.0950).Similarly, removing TS‑Align leads to performance declines. On Industrial, HR@5 falls by 3.64\% (from 0.1153 to 0.1111) and NDCG@5 by 0.21\% (from 0.0930 to 0.0928), while on Office, HR@5 decreases by 0.53\% (from 0.1307 to 0.1300) and NDCG@5 by 1.45\% (from 0.1100 to 0.1084).% 
% These results indicate that:
% (a) SA‑Init provides a strong initialization for token embeddings, facilitating more effective optimization during post-training.
% (b) TS‑Align injects fine-grained semantic coherence into SID representation, boosting downstream ranking accuracy.
% (2) \textbf{The combination of SA-Init and TS-Align yields the strongest SFT-based optimization.} This synergy suggests that SA-Init and TS-Align are complementary components, as each addresses different aspects of model optimization, and together they drive maximal performance gains.

% % 1.94\%, 2.04\%, 8.33\%, 13.36\%
% 3.64\%, 0.21\%, 0.53\%, 1.45\%% 

\subsubsection{Impact of Token Initialization Depth.}
Although the preceding results demonstrate the effectiveness of SA-Init, the underlying mechanisms through which it influences training remain unclear. We therefore conduct a set of controlled experiments to investigate the aspects of training affected by SA-Init.

Specifically, we compare the following four initialization strategies for the SID token embeddings:
\begin{itemize}[leftmargin=*]
\item \textit{Random Init}: Initialize SID token embeddings using Qwen2.5's default initialization method.

\item \textit{SA-Init 1}: Apply SA-Init to initialize the first SID level of the SID sequence's token embeddings.

\item \textit{SA-Init 2}: Apply SA-Init to initialize the first two SID levels of the SID sequence's token embeddings.

\item \textit{SA-Init 3}: Apply SA-Init to initialize all three SID levels of the SID sequence's token embeddings.

\end{itemize}

We track training dynamics on the Industrial, Office, and Toys datasets, recording the evaluation loss and HR@20 throughout training. As illustrated in Figure~\ref{fig:hr_and_loss}, our main observations are:
(1) \textbf{Faster optimization.} Across all three datasets, SA-Init generally reduces early-stage evaluation loss and accelerates the improvement of HR@20 compared with Random Init. This suggests that semantically grounded initialization provides a more informative starting point, reducing the burden of learning SID token semantics from scratch during SFT.
(2) \textbf{Benefits of deeper initialization.} Initializing more SID levels generally yields faster convergence and stronger HR@20, with SA-Init 3 achieving the best or near-best performance throughout training. For example, at the first evaluation point on Toys, SA-Init 3 already reaches approximately 0.10 HR@20, whereas Random Init remains close to zero.
\subsection{\ours as a Backbone for Reinforcement Learning (RQ3)}

Recent breakthroughs in large language models, exemplified by the DeepSeek series~\citep{guo2025deepseek}, have motivated the recommendation community to adopt the SFT-then-RL paradigm~\citep{DBLP:journals/corr/abs-2510-24431}. In this workflow, supervised fine-tuning (SFT) establishes the initial policy, while reinforcement learning (RL) further aligns the policy with user preferences and ranking metrics. However, RL optimization is sensitive to policy initialization; a backbone that lacks a strong understanding of the semantic identifier space may produce unstable reward signals and suboptimal convergence.

% Recent generative recommenders commonly adopt an SFT-then-RL pipeline~\citep{DBLP:journals/corr/abs-2510-24431}. Since RL optimization is sensitive to its initial policy, we investigate whether the semantic grounding provided by \ours can serve as a stronger SFT backbone.

\begin{table*}[t]
\caption{Performance comparison between MiniOneRec and MiniOneRec* on the \textit{Industrial} and \textit{Toys} datasets. MiniOneRec* is obtained by replacing the original SFT-based backbone in MiniOneRec with \ours. HR and N denote Hit Ratio and NDCG, respectively.}
% \vspace{-0.1in}
\label{minionerec}
\centering
\setlength{\tabcolsep}{1mm}{
    \resizebox{0.93\textwidth}{!} {
        \begin{tabular}{ll cccc cccc cccc}
        \toprule

        \multirow{2.4}{*}{Models} & \multicolumn{6}{c}{\textbf{Industrial}} & \multicolumn{6}{c}{\textbf{Toys}} \\
        \cmidrule(lr){2-7} \cmidrule(lr){8-13} & HR@3 & N@3 & HR@5 & N@5 & HR@10 & N@10 & HR@3 & N@3 & HR@5 & N@5 & HR@10 & N@10 \\
        \midrule

        MiniOneRec & 0.0991 & 0.0860 & 0.1160 & 0.0928 & 0.1443 & 0.1020 & 0.0652 & 0.0539 & 0.0802 & 0.0600 & 0.1054 & 0.0682 \\
        \textbf{MiniOneRec*} & \textbf{0.1118} & \textbf{0.0995} & \textbf{0.1295} & \textbf{0.1068} & \textbf{0.1568} & \textbf{0.1156} & \textbf{0.0695} & \textbf{0.0582} & \textbf{0.0852} & \textbf{0.0647} & \textbf{0.1107} & \textbf{0.0729} \\
        
        \midrule
        Improvement & \textcolor{ourcolor1}{+12.81\%} & \textcolor{ourcolor1}{+15.68\%} & \textcolor{ourcolor1}{+11.63\%} & \textcolor{ourcolor1}{+15.08\%} & \textcolor{ourcolor1}{+8.66\%} & \textcolor{ourcolor1}{+13.44\%} & \textcolor{ourcolor1}{+6.59\%} & \textcolor{ourcolor1}{+7.97\%} & \textcolor{ourcolor1}{+6.23\%} & \textcolor{ourcolor1}{+7.83\%} & \textcolor{ourcolor1}{+5.02\%} & \textcolor{ourcolor1}{+6.89\%} \\
        \bottomrule
        \end{tabular}
    }
}
% \vspace{-0.10in}
\end{table*}

% \begin{table}[t]
% \setlength{\tabcolsep}{3pt}
% \caption{Performance comparison between MiniOneRec and MiniOneRec* on the \textit{Industrial} \fjw{and \textit{Toys}} dataset. MiniOneRec* is obtained by replacing the original SFT-based backbone in MiniOneRec with \ours.}
% \label{minionerec}
% % \vspace{-5pt}
% \small
% \begin{tabular}{llccc}
% \toprule
% \multicolumn{1}{c}{\multirow{1}{*}{\textbf{Dataset}}} & \textbf{Metric} & \textbf{MiniOneRec} & \textbf{MiniOneRec*} & Improvement \\ \midrule
% % \rowcolor{gray!20} 
% \multirow{6}{*}{\textbf{Industrial}} & HR@3 & 0.0991 & \textbf{0.1118} & 12.81\% \\
%  & HR@5 & 0.1160 & \textbf{0.1295} & 11.63\% \\
%  & HR@10 & 0.1443 & \textbf{0.1568} & 8.66\% \\
%  & NDCG@3 & 0.0860 & \textbf{0.0995} & 15.68\% \\
%  & NDCG@5 & 0.0928 & \textbf{0.1068} & 15.08\% \\
%  & NDCG@10 & 0.1020 & \textbf{0.1156} & 13.44\% \\
%  \midrule
% % \rowcolor{gray!20} 
% \multirow{6}{*}{\textbf{Toys}} & HR@3 & 0.0652 & \textbf{0.0695} & 6.59\% \\
%  & HR@5 & 0.0802 & \textbf{0.0852} & 6.23\% \\
%  & HR@10 & 0.1054 & \textbf{0.1107} & 5.02\% \\
%  & NDCG@3 & 0.0539 & \textbf{0.0582} & 7.97\% \\
%  & NDCG@5 & 0.0600 & \textbf{0.0647} & 7.83\% \\
%  & NDCG@10 & 0.0682 & \textbf{0.0729} & 6.89\% \\
% \bottomrule
% \end{tabular}
% \vspace{-5pt}
% \end{table}

To evaluate whether the fine-grained semantic grounding of \ours can facilitate more effective RL optimization, we integrate it into \textbf{MiniOneRec}~\cite{DBLP:journals/corr/abs-2510-24431}, a state-of-the-art open-source framework for scaling generative recommendation. Specifically, we replace MiniOneRec's original SFT backbone with \ours (denoted as \textbf{MiniOneRec*}) while keeping the subsequent RL (e.g., GRPO) configurations identical.
The results on the \textit{Industrial} and \textit{Toys} datasets are summarized in Table~\ref{minionerec}, yielding several critical insights:
(1) \textbf{Consistent improvements after RL.} MiniOneRec* outperforms MiniOneRec across all metrics, with relative gains of 8.66\%--15.68\% on Industrial and 5.02\%--7.97\% on Toys. In particular, NDCG@3 and NDCG@5 improve by 15.68\% and 15.08\% on Industrial, respectively. The consistent gains on both datasets demonstrate that the fine-grained semantic grounding learned during SFT remains beneficial after subsequent RL optimization.
(2) \textbf{Stronger ranking quality.} On both datasets, the relative NDCG gains are consistently larger than the corresponding HR gains at the same cutoff. For example, at $K=3$, MiniOneRec* improves NDCG by 15.68\% versus 12.81\% in HR on Industrial, and by 7.97\% versus 6.59\% on Toys. This result indicates that \ours not only improves target-item retrieval but also helps the RL-enhanced model rank the target item more accurately, supporting its role as a stronger initial policy.
% (1) \textbf{Consistent improvements across ranking metrics.} MiniOneRec* consistently outperforms the original MiniOneRec on both datasets and across all metrics. On Industrial, the relative improvements range from 8.66\% to 15.68\%, while on Toys they range from 5.02\% to 7.97\%. These results demonstrate that the fine-grained semantic grounding provided by \ours establishes a stronger SFT policy for subsequent RL optimization across datasets of different scales and sparsity levels.
% (2) \textbf{\ours as a high-quality policy prior.} The superiority of MiniOneRec* demonstrates that \ours provides a stronger starting point for RL. Unlike random initialization, which requires RL to learn token semantics and user preferences simultaneously, SA-Init and TS-Align separate these challenges. By grounding SID tokens in the LLM's language space beforehand, \ours reduces the search space and allows RL to focus on optimizing recommendation rewards.
% (3) \textbf{Enhanced exploration efficiency.} On both datasets, the relative gains in NDCG are consistently larger than the corresponding gains in Hit Ratio at the same cutoff. For example, at $K=3$, MiniOneRec* improves NDCG by 15.68\% versus 12.81\% in Hit Ratio on Industrial, and by 7.97\% versus 6.59\% on Toys. This indicates that \ours helps the model not only retrieve the target item but also rank it more precisely. One possible explanation is that fine-grained token-level alignment provides a more coherent SID representation space, allowing the RL policy to explore more effectively.

In summary, the transition from SFT to RL is not a disjoint process; the quality of semantic integration during SFT influences the effectiveness of subsequent RL optimization. By creating a semantically coherent representation space, \ours provides a strong backbone for RL-enhanced LLM-based recommendation.

\begin{figure}[htbp]
\centering
\includegraphics[width=0.48\textwidth]{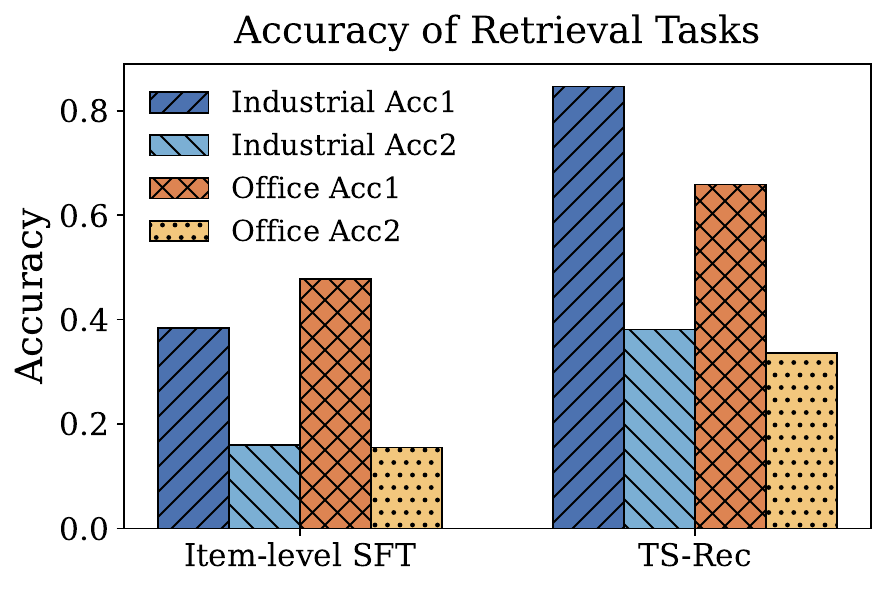}
% \vspace{-20pt}
\caption{Accuracy on the generative retrieval tasks for the Industrial and Office datasets. ACC1 and ACC2 denote the exact-match accuracy of Title2SID and SID2Title, respectively.}
\label{fig:sid_retrieval}
\end{figure}
\vspace{-20pt}

% \subsection{The effect of \ours on SID Comprehension (RQ4)}
\subsection{SID Comprehension Analysis (RQ4)}

% To evaluate whether \ours can effectively capture and utilize fine-grained SID semantics, we design two generative retrieval tasks:
% \begin{itemize}[leftmargin=*]
% \item \textit{Title2SID task:} Given the textual title of an item, predict its corresponding SID token sequence.
% \item \textit{SID2Title task:} Given the SID token sequence of an item, predict its corresponding textual title.
% \end{itemize}

To evaluate whether \ours can effectively capture and utilize fine-grained SID semantics, we design two generative retrieval tasks: \textit{Title2SID}, which predicts an item's SID token sequence from its title, and \textit{SID2Title}, which reconstructs the corresponding item title from its complete SID sequence.

Both tasks use beam-search decoding with a beam width of 5. We compare \ours with random initialization combined with item-level SFT on the Industrial and Office datasets. The results in Figure~\ref{fig:sid_retrieval} yield two observations:
(1) \textbf{\ours substantially improves both retrieval tasks.} On Industrial, ACC1 and ACC2 improve by 120.89\% and 138.34\%, respectively; on Office, the corresponding gains are 37.78\% and 116.94\%. These consistent improvements demonstrate that \ours captures and utilizes SID-level semantics more effectively than item-level SFT.
% (2) \textbf{\ours strengthens the connection between language and SIDs.} Its performance on both mapping directions demonstrates a stronger correspondence between natural-language descriptions and structured semantic identifiers, which may also benefit semantic retrieval and cold-start recommendation.
% (2) \textbf{The improvement is particularly pronounced for SID2Title.} The relative gain in ACC2 is larger than that in ACC1 on both datasets, indicating that token-level semantic alignment is especially helpful for recovering language information from structured SIDs. Nevertheless, the absolute ACC2 remains lower than ACC1, likely because generating an exact item title involves a larger output space and greater lexical variability than generating a constrained SID sequence.
% (3) \textbf{\ours strengthens the connection between language and SIDs.} The consistent improvements in both mapping directions demonstrate a stronger correspondence between natural-language descriptions and structured semantic identifiers. This bidirectional understanding may also benefit semantic retrieval and cold-start recommendation.
(2) \textbf{\ours strengthens the connection between language and SIDs.} The relative gain in ACC2 is larger than that in ACC1 on both datasets, showing that token-level alignment is particularly helpful for recovering language information from SIDs. The improved bidirectional correspondence may also benefit semantic retrieval and cold-start recommendation.

\section{Conclusion and Future Work}

In this work, we introduced \ours, a framework that integrates fine-grained SID semantics into LLM-based generative recommendation. To address semantically meaningless initialization and coarse-grained alignment, SA-Init provides SID tokens with cluster-level semantic priors, while TS-Align establishes token-level mappings between the SID and language spaces. Experiments on three real-world benchmarks show that \ours consistently outperforms traditional and generative baselines and provides a stronger SFT backbone for subsequent RL optimization. Ablation studies confirm the complementary effects of both components, and the initialization-depth analysis shows that deeper semantic initialization generally accelerates convergence and improves recommendation performance. Moreover, \ours strengthens bidirectional mappings between SIDs and natural-language descriptions, which may also benefit cold-start recommendation.

% 未来工作

While \ours integrates fine-grained semantics through embedding initialization and SFT-based alignment, how to preserve and further refine token-level semantics during RL remains underexplored. Future work will investigate semantic reward functions that explicitly evaluate the consistency between generated SID tokens and their cluster-level meanings. Such rewards may discourage semantically inconsistent generations while improving the alignment between recommendation policies and user preferences. We also plan to extend \ours to multimodal recommendation by extracting shared token-level semantics from textual, visual, and behavioral item information.

% \clearpage

\balance
\bibliographystyle{ACM-Reference-Format}
\bibliography{reference}

% \conferenceonly{
%     \newpage
%     \appendix
%     \input{sections/appendix}
%     \label{sup}
% }

\newpage
\appendix
\onecolumn

\label{appendix}

\section{Appendix}

\subsection{Prompt}\label{appendix:prompt}

\begin{tcolorbox}[
colback=gray!5!white,
colframe=gray!60!black,
title=\textbf{ Prompt for Semantic Extraction},
fonttitle=\bfseries,
breakable,
boxrule=0.8pt,
left=2pt, right=2pt, top=2pt, bottom=2pt
]
\textbf{\# Role:}

You are a Senior Product Content Analyst and Catalog Specialist with expertise in {\color{gray}\{category\}} product classification and summarization.

\textbf{\# Task:}
Analyze a provided list of diverse product items and synthesize their shared characteristics into a single, cohesive "Product Family Description," then produce a related keyword list.

\textbf{\#\#Input Format:} Each item provides:
\begin{itemize}
    \item Title: product name
    \item Description: detailed text about features, functions, and specifications
    \item Brand: brand name
    \item Categories: classification field(s)
\end{itemize}

\textbf{\#\# Core Objective:}

\begin{itemize}
    \item Identify Commonalities: Determine the underlying functional, design, usage, or performance traits that ALL listed items share.
    \item Filter Fluff: Exclude brand-specific marketing slogans, proprietary technology names, or promotional phrases (e.g., "award-winning," "exclusive edition").
    \item Abstract Specifications: Replace exact measurements, quantities, or configurations with generalized terms (e.g., "various sizes," "multiple capacity options," "bulk packaging").
    \item Synthesize: Produce a high-level overview that describes the product family as a category, rather than a list of individual items.
    \item Extract Keywords: Identify 10-20 domain-relevant, semantically meaningful keywords derived from the description.
\end{itemize}

\textbf{\#\# Style \& Tone:}
\begin{itemize}
    \item Professional, neutral, and domain-agnostic.
    \item Objective third-person, suitable for product catalogs and technical documentation.
    \item One paragraph of 3--5 sentences that flows naturally.
    \item Clear, concise, and focused on function, durability, applicability, or end-user benefits.
\end{itemize}

\textbf{\#\# Constraints:}
\begin{itemize}
    \item No brand names in the output.
    \item No bullet points; provide a cohesive, continuous paragraph.
    \item Ensure the description is broad enough to encompass all items in the list but specific enough to convey meaningful shared traits.
    \item Avoid overly promotional adjectives; emphasize factual, verifiable characteristics.
\end{itemize}

\textbf{\# Input:}

Items:

{\color{gray}\{items\}}

\end{tcolorbox}

\subsection{Fine-tuning Task for Recommendation}\label{appendix:sft_tasks}

\subsubsection{Sequential Item Prediction}

\begin{tcolorbox}[
breakable,
boxrule=0.8pt,
left=2pt, right=2pt, top=2pt, bottom=2pt
]
\textbf{Instruction ($x$):}

The user has interacted with items <a\_251><b\_198><c\_114>, ..., <a\_251><b\_198><c\_132> in chronological order. Can you predict the next possible item that the user may expect?

\textbf{Response ($y$):}

<a\_251><b\_198><c\_238>
\end{tcolorbox}

\subsubsection{Explicit SID-Language Alignment}

\begin{tcolorbox}[
breakable,
boxrule=0.8pt,
left=2pt, right=2pt, top=2pt, bottom=2pt
]
\textbf{Instruction ($x$):}

Which item has the title: 6061 Aluminum Round Rod, Unpolished (Mill) Finish, Extruded, T6511 Temper, ASTM B221, 2\" Diameter, 12\" Length?

\textbf{Response ($y$):}

<a\_251><b\_198><c\_238>
\end{tcolorbox}

\begin{tcolorbox}[
breakable,
boxrule=0.8pt,
left=2pt, right=2pt, top=2pt, bottom=2pt
]
\textbf{Instruction ($x$):}

What is the title of item "<a\_251><b\_198><c\_238>"?

\textbf{Response ($y$):}

6061 Aluminum Round Rod, Unpolished (Mill) Finish, Extruded, T6511 Temper, ASTM B221, 2\" Diameter, 12\" Length
\end{tcolorbox}

\subsubsection{Asymmetric Item Prediction-1}

\begin{tcolorbox}[
breakable,
boxrule=0.8pt,
left=2pt, right=2pt, top=2pt, bottom=2pt
]
\textbf{Instruction ($x$):}

The user has sequentially interacted with items {history}. Can you recommend the next item for him? Tell me the title of the item.

\textbf{Response ($y$):}

6061 Aluminum Round Rod, Unpolished (Mill) Finish, Extruded, T6511 Temper, ASTM B221, 2\" Diameter, 12\" Length
\end{tcolorbox}

\subsubsection{Asymmetric Item Prediction-2}

\begin{tcolorbox}[
breakable,
boxrule=0.8pt,
left=2pt, right=2pt, top=2pt, bottom=2pt
]
\textbf{Instruction ($x$):}

Please review the user's historical interactions: <a\_251><b\_198><c\_114>, ..., <a\_251><b\_198><c\_132>, and describe what kind of item he still needs.

\textbf{Response ($y$):}

The 6061 aluminum round rod with T6511 temper has an unpolished (mill) surface, ..., while weldability characterizes the ability to be welded.
\end{tcolorbox}

\end{document}